\documentclass[10pt,letterpaper,twocolumn]{article} 

\newcommand{\be}{\begin{equation}}
\newcommand{\ee}{\end{equation}}
\newcommand{\bd}{\begin{displaymath}}
\newcommand{\ed}{\end{displaymath}}

\newcommand{\bea}{\begin{eqnarray}}
\newcommand{\eea}{\end{eqnarray}}

\renewcommand{\phi}{\varphi}
\renewcommand{\vec}[1]{{\mathbf #1}}

\usepackage{ol2}
\usepackage[draft]{hyperref}
\usepackage{amsmath}

\begin{document}

\twocolumn[ 

\title{Resonant Bend Loss in Leakage Channel Fibers}


\author{R.~A.~Barankov$^{1,*}$, K.~Wei$^2$, B.~Samson$^2$,
S.~Ramachandran$^{1,*}$}

\address{
$^1$Photonics Center, Department of Electrical and Computer Engineering, \\
Boston University, 8 Saint Mary's Street, Boston, Massachusetts 02215, USA\\
$^2$Nufern, East Granby, Connecticut 06026, USA \\ $^*$Corresponding authors:
barankov@bu.edu, sidr@bu.edu
}

\begin{abstract}

Leakage channel fibers, designed to suppress higher-order modes, demonstrate
resonant power loss at certain critical radii of curvature. Outside the
resonance, the power recovers to the levels offset by the usual mechanism of
bend-induced loss. Using C$^2$-imaging, we experimentally characterize this
anomaly and identify the corresponding physical mechanism as the radiative
decay of the fundamental mode mediated by the resonant coupling to a cladding
mode.

\end{abstract}
] 

Resonantly enhanced leakage-channel fibers (LCFs) with large mode areas are
designed to provide high-power propagation of diffraction-limited beams in
high-power fiber lasers~\cite{Dong2009}. The microstructure of these fibers is
tailored to enhance the loss of higher-order modes (HOMs) while maintaining
tolerable loss of the fundamental mode, resulting in single-mode operation
with large field diameters. In the existing designs~\cite{Dong2009}, the
index-continuity at the boundary in the azimuthal direction is broken. As a
result, all modes propagating in the fiber are coupled to the radiative modes.
However, the presence of gaps at the interface naturally promotes significant
differential power-loss dominated by HOMs~\cite{Dong2009}.

\begin{figure}[!htb]
\centering
\includegraphics[width=7.5cm]{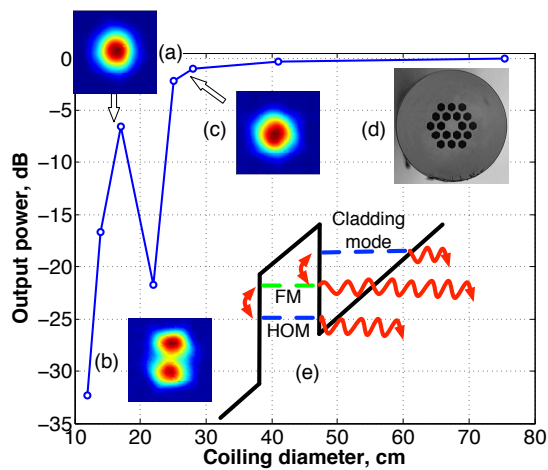}
\caption{(color online) Output power as a function of coiling diameter. {\it Insets} (a-c):
Output mode profiles at different coiling diameters; (d) Cross-section of the
LCF; (e) Power loss mechanisms in bent fibers}
\label{fig:LCF_power}
\end{figure}

The physical mechanism underlying the LCF design suggests sensitivity of light
propagation to the coiling conditions of the fibers. Coiled LCFs experience
additional power loss resulting from the usual bend-induced coupling of the
guided modes to the radiative
modes~\cite{Marcuse1976,Wong2005,Dong2006,Saitoh2011}. This mechanism is
illustrated in Fig.~\ref{fig:LCF_power}(e), where the core modes propagating
in the slanted index-profile, representing the effect of fiber-bending,
directly couple to the radiative modes. Certainly, the loss is significantly
higher for HOMs than for the fundamental mode, as dictated by a narrower
tunneling barrier.

This bend-loss becomes significant for small enough coiling
radii~\cite{Love1989}. The critical radius has been estimated, in the case of
photonic-crystal fibers~\cite{Birks1997}, as $R_c\sim \Lambda^3/\lambda^2$ in
the short-wavelength limit, where $\lambda$ is the wavelength of light, and
$\Lambda$ is the characteristic core size~\cite{Birks1997,Nielsen2004}. This
estimate should also hold for leakage-channel fibers characterized by similar
geometry. For example, for the LCFs with core size $\Lambda\sim 50\, {\rm\mu
m}$, significant bend loss is expected at $R_c\sim 10 \, {\rm cm}$ in the $1\,
{\rm \mu m}$ spectral range.

In addition, bending of fibers induces intermodal coupling of the modes. The
coupling increases dramatically when the effective indices of the two modes
approach one another as a function of the fiber curvature as shown in
Fig.~\ref{fig:LCF_power}(e). The anti-crossing occurs at some critical radius,
which, in general, depends on the properties of the coupled
modes~\cite{Love2007,Yao2009}. Specifically, the fundamental mode can couple
to a leaky HOM of the core, or to a quasi-guided cladding mode, as described
in~\cite{Murakami1978,Renner1992}, which leads to strong attenuation of light
power.

In this work, we explore the interplay of these effects in a large-mode area
LCF and observe an enhanced power loss at a certain coiling diameter. The
results of power measurements in the $1$-${\rm \mu m}$ spectral range are
shown in Fig.~\ref{fig:LCF_power}. Specifically, we find that for non-critical
coiling radii, light propagation is dominated by the fundamental mode, as
output near-field images in Fig.~\ref{fig:LCF_power}(a) and (c) illustrate.
Thus, at non-critical radii the power-loss can be explained by the usual
bend-loss mechanisms~\cite{Marcuse1976,Birks1997}. In contrast, at the
critical coiling radius, a higher-order mode dominates light propagation as
shown in Fig.~\ref{fig:LCF_power}(b). While simple bend-loss measurements in
Fig.~\ref{fig:LCF_power} reveal this anomalous behavior, they do not identify
the physical mechanism. We characterize the observed anomaly by the
C$^2$-imaging method~\cite{Schimpf2011} and find that the effect is dominated
by the resonant mode coupling of the fundamental mode to a quasi-guided
cladding mode~\cite{Murakami1978,Renner1992}. As a result, this method
provides critical feedback for future fiber-designs, in which the critical
radius $R_c$ should be defined for specific amplifier packaging constraints.

\begin{figure}[!htb]
\centering
\includegraphics[width=8cm]{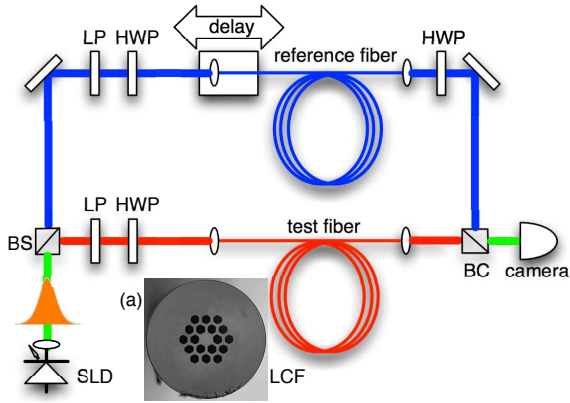}
\caption{(color online) C$^2$-imaging setup: SLD -- superluminescent diode, LP -- linear
polarizer, HWP -- half-wave plate, BS --
beam-splitter, BC -- beam combiner. {\it Inset} (a): Cross-section of the LCF}
\label{fig:LCF_setup}
\end{figure}

The tested LCF of $285$-${\rm cm}$-length has a core diameter of $50\,{\rm\mu
m}$ and cladding diameter of $400\,{\rm\mu m}$. The two rings of low-index
(fluorine-doped) silica regions shown in Fig.~\ref{fig:LCF_setup}(a) provide
the leakage channel. The core, made of silica, is index matched to the outer
silica glass. A high-index regular acrylate coating applied to the cladding
ensures stripping of the cladding modes. The LCF has been designed to have
negligible HOM content at lengths greater than $3\,{\rm m}$. The input end of
the fiber was spliced to a single-mode fiber to provide the same in-coupling
conditions throughout the experiments.

The modal content of the LCF was analyzed using the C$^2$-imaging
method~\cite{Schimpf2011} modified to account for elliptical polarization of
the test beam. The basic idea of the method is to study the interference of
the test beam with an external reference beam, and detect different waveguide
modes in time-domain by changing the relative optical paths of the two
beams.

Figure~\ref{fig:LCF_setup} shows a schematic diagram of the experimental setup
based on a Mach-Zehnder interferometer. We used a superluminescent diode (SLD)
source centered at $\lambda\sim 1050\,{\rm nm}$ with a spectral width of $\sim
30\, {\rm nm}$. The output beam of the LCF (focused at the imaging plane)
interferes with the collimated reference beam radiated from the reference
fiber. The length of polarization-maintaining reference fiber is chosen to
compensate for the optical path difference between the two paths and also to
reduce the effects of group-velocity dispersion of the LCF. The latter is
important since we employ a light source with a relatively broad spectrum,
which, in the absence of dispersion compensation, leads to significant
dispersive broadening of the cross-correlation signal. In particular, the
mode-specific resolution of the C$^2$-imaging method is defined by the
spectral width of the source and also by the dispersion mismatch of the
reference and test modes. In large-mode-area LCFs, the material dispersion
dominates spectral broadening, which results in similar dispersion
properties of HOMs. We chose the length of the reference fiber to match the
material dispersion of LCF and, thus, reduced the dispersive broadening of all
the modes simultaneously.

The cross-correlation signal ${\cal P}(\vec r, \tau)$ as a function of the
group delay $\tau$ and the coordinate $\vec r$ in the imaging plane is
detected by the camera for different delay-stage positions $d=c\tau$ ($c$ is
the speed of light in vacuum):
\be\label{eq:cross_correlation_local}
{\cal P}(\vec r, \tau)=\sum_m p_m{\cal G}_{mr}^2(\tau-\tau_{mr}) I_{m}(\vec
r).
\ee
Here, the summation extends over the modes propagating in LCF, $p_m$ is the
relative modal power of $m$-th mode ($\sum_m p_m=1$), ${\cal G}_{mr}(\tau)$ is
the mutual coherence function of the reference and test beams, $I_{m}(\vec
r)$ is the modal intensity, and $\tau_{mr}$ is the relative group delay of the
$m$-th mode with respect to the reference mode. 

Coiling of the fiber affects polarization properties of the test beam. To
extract the power of every elliptically-polarized mode, we have recorded the
cross-correlation trace for two orthogonal polarization states of the
reference beam. The resulting trace, combining the two measurements, is
represented in Eq.~(\ref{eq:cross_correlation_local}). The intensity
distribution $I_m(\vec r)$ of every mode was obtained by integrating this
expression over the time extent of the mode. The relative power of the modes
is encoded in the net cross-correlation trace obtained by integration of the
spatially-dependent trace in Eq.~(\ref{eq:cross_correlation_local}) over the
imaging plane position.

\begin{figure}[!htb]
\centering
\includegraphics[width=8cm]{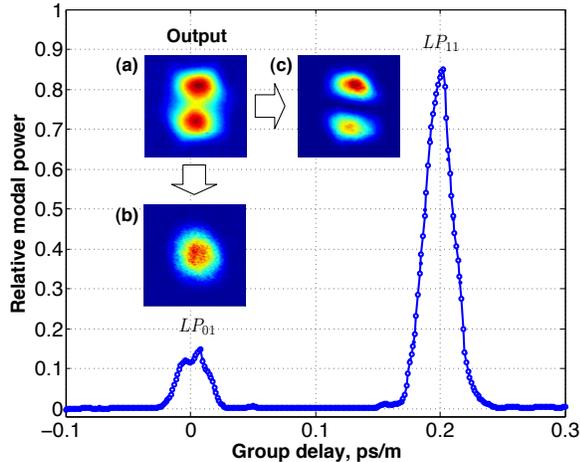}
\caption{(color online) Relative modal power as a function of group delay at the resonance
($R_c=11\,{\rm cm}$). {\it Insets} (a-c): output image and the images of
reconstructed modes}
\label{fig:LCF_resonance}
\end{figure}

The result of this procedure, applied to the cross-correlation traces recorded
at the critical coiling radius, is shown in Fig.~\ref{fig:LCF_resonance}. In
this figure, the cross-correlation peaks identify the modes propagating in the
fiber at the corresponding relative group delays. The shape of the peaks
reflects the corresponding mutual coherence functions, while the peak values
encode the relative power of the modes. The insets demonstrate the intensities
of the reconstructed modes.

The dependence of the output power on the coiling diameter was measured using
a power meter (as shown earlier in Fig.~\ref{fig:LCF_power}). Interestingly,
we observe a dramatic decrease of the output light power at a specific coiling
diameter $R_{exp}\approx 11 \,{\rm cm}$, which is close to the estimate
$R_c\sim 10 \,{\rm cm}$ for the onset of substantial power-loss. At this
resonant coiling condition, the output image demonstrates domination of HOM’s
as shown in Fig.~\ref{fig:LCF_power}(b), while the output images in
Fig.~\ref{fig:LCF_power}(a) and~\ref{fig:LCF_power}(c), recorded at coiling
diameters outside the resonance, indicate single-mode operation. These
measurements, however, do not indicate which of the two probable mechanisms,
-- the intermodal coupling of the leaky core modes or the coupling between a
core mode and quasi-guided cladding mode, -- is responsible for the observed
resonant behavior.

\begin{figure}[!htb]
\centering
\includegraphics[width=7.5cm]{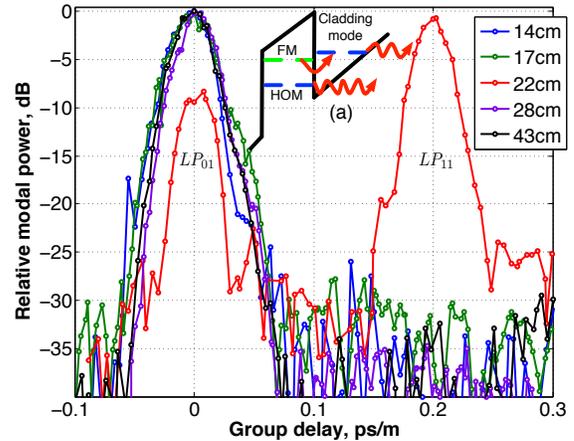}
\caption{(color online) Relative modal power of $LP_{01}$ and $LP_{11}$ (peak values) as a
function of the relative group delay, for different coiling diameters. {\it
Inset} (a): Bend-induced coupling of the FM and cladding mode as a mechanism
of power-loss}
\label{fig:LCF_out_resonance}
\end{figure}

C$^2$-imaging~\cite{Schimpf2011} provides insight into the resonant behavior.
The envelope of the integrated correlation trace shown in
Fig.~\ref{fig:LCF_resonance} at the critical coiling diameter demonstrates two
modes ($LP_{01}$ and $LP_{11}$) propagating with a relative group delay of
about $0.2\, {\rm ps/m}$, with the fundamental mode contributing only about
15\% of the total power. In contrast, at other coiling diameters, HOM’s are
suppressed at power levels below $-25\,{\rm dB}$, as shown in
Fig.~\ref{fig:LCF_out_resonance}. The well-defined group delay between the
strongly attenuated fundamental mode and HOM suggests that the former
resonantly couples to a cladding mode, as shown in
Fig.~\ref{fig:LCF_out_resonance}(a). In the alternative mechanism, coupling
between the core modes would lead to the appearance of a broad feature in
C$^2$ trace due to the distributed nature of the coupling, and thus can not be
detected by this method. However, the mechanism, if present, may affect the
observed modal power distribution. In a set of similar measurements conducted
on an LCF of smaller length of about $180\,{\rm cm}$, a relatively shallow
resonance was found at the same coiling diameter, with the relative group
delay of $\sim 0.2\, {\rm ps/m}$ between $LP_{01}$ and $LP_{11}$ modes.
Likewise, we expect significantly deeper resonances for longer fiber lengths.

In summary, coiled few-mode fibers experience bend-induced coupling between
the core and cladding modes as well as power-loss via direct coupling of core
modes to the radiative modes. Using C$^2$-imaging, we explore the interplay
between these phenomena in LCFs. We identify a resonant power-loss mechanism,
in which a cladding mode mediates the coupling of the fundamental mode to
the radiative modes. The effect becomes evident at a specific coiling
diameter, where we observe a dramatic decrease of the output power. Outside
the resonance, the power recovers to the levels determined by the usual
bend-loss mechanism. We expect that quantitative characterization of this type
would be critical in fiber laser applications using LCFs, since allowed
coiling diameters would directly impact packaging conditions of high-power
fiber-lasers.

This work was done with the support of ARL Grant No. W911NF-06-2-0040 and ONR
grant nos. N00014-11-1-0133 \& N00014-11-1-0098.

\end{document}